\documentstyle[aps,manuscript,graphicx]{revtex}

\begin{document}

\title{Two-surface wave decay}

\author{Andrea Macchi\thanks{E-mail: {\tt macchi@df.unipi.it}}, 
Fulvio Cornolti and Francesco Pegoraro}
\address{Dipartimento di Fisica, Universit\'{a} di Pisa e \\ 
Istituto Nazionale Fisica della Materia (INFM), sezione A, \\
Piazza Torricelli 2, I-56100 Pisa, Italy}

\maketitle
\begin{abstract}
The parametric excitation of 
pairs of electron surface waves (ESW) 
in the interaction of an ultrashort, intense laser pulse
with an overdense plasma
is discussed using an analytical model.
The plasma has a simple step-like density profile.
The ESWs can be excited either by the electric or by
the magnetic part of the Lorentz force exerted by the laser 
and, correspondingly, have frequencies around $\omega/2$ 
or $\omega$, where $\omega$ is the laser frequency.
\end{abstract}


\section{Introduction}

Surface wave (SW) excitation has been widely studied 
in the past as a mechanism for electromagnetic (EM) 
wave absorption in highly inhomogeneous plasmas 
in several regimes and for 
different target geometries
\cite{kindel,dragila} and, more recently, in the 
specific case of solid 
targets irradiated by intense, ultrashort laser pulses
\cite{kupersztych,gamaly1}.
The key problem in the linear mode conversion of the 
EM wave (the laser pulse) into an SW
of the same frequency $\omega$ 
is that the phase matching between the two waves
is not possible at a simple plasma-vacuum interface.
The reason for this is that for an SW vave the wavevector along
the plasma surface $k_s$ is larger than $\omega/c$, 
and thus cannot be equal 
to the component of the EM wavevector 
in the same direction $k_t=(\omega/c)\sin\theta$, where
$\theta$ is the incidence angle of the laser pulse.
Linear mode conversion of the laser pulse 
into an SW is possible in specifically tailored density
profiles, e.g. for a ``double-step''
density profile (i.e. at the interface between an
underdense plasma region where $\omega>\omega_p$
and an overdense region where $\omega<\omega_p$,
with $\omega_p$ the plasma frequency),
or if a periodic surface modulation 
with wavevector $k_p=k_t-k_s$ exists, 
such that the matching condition for the wavevectors is
\begin{equation}
k_t=k_s+k_p .
\end{equation}
In practical applications, corrugated targets are used to 
optimize SW generation at a given incidence angle.
Other possible schemes of SW excitation are discussed,
e.g., in Refs.\cite{dragila,kupersztych}.

In this paper, we discuss a different mechanism of 
electron surface wave (ESW) excitation by ultrashort 
laser pulses, based on the 
generation of {\em two} counterpropagating ESWs.
The basic idea is as follows. An intense laser 
pulse impinging on an overdense, 
step-boundary plasma drives electron oscillations at the 
frequencies $\omega$ and $2\omega$ by the electric and the 
magnetic part of the Lorentz force , respectively. 
The electron response may be viewed as a superposition
of ``forced'' surface modes with frequency $\omega_0=\omega$ or 
$2\omega$, respectively, and wavevector 
$k_0=(\omega_0/c)\sin\theta$.
Each of these forced modes may ``decay'' parametrically into two 
ESWs (labeled ``$+$'' and ``$-$'', respectively)
provided that 
the matching conditions for a three-wave process hold:
\begin{eqnarray}
k_0=k_{+}+k_{-},\\
\omega_0=\omega_{+}+\omega_{-}.
\label{eq:MC}
\end{eqnarray}
No pre-imposed target modulation is required to 
satisfy these relations, as depicted 
in Fig.\ref{fig:ESW}. The ESW frequencies 
may be written as 
$\omega_{\pm}=\omega_{0}/2 \pm \delta \omega$.
Thus, the ``decay'' of the mode excited by the electric
(magnetic) force leads to the generations of ESWs
with frequencies around $\omega/2$ ($\omega$). 
We call this process ``two-surface wave decay'' (TSWD)
in analogy to the well-known ``two-plasmon 
decay'' parametric instability in laser-plasma
interactions. 

Earlier investigations of parametric and three-wave
nonlinear processes involving ESWs have been previously
carried out in different regimes \cite{aliev,stenflo}.
In particular, the parametric excitation of pairs of subharmonic ESWs in a magnetized plasma by an external EM wave was studied in Refs.\cite{gradov1}.
These studies were restricted to the electrostatic limit in which the ESW
frequency is $\omega_{s} \simeq \omega_p/\sqrt{2}$ 
(see \cite{kaw} and eq.(\ref{eq:SW_DR}) below)
and does not depend on the wavevector $k_{s}$ in a semi-infinite,
step boundary plasma 
(e.g. with a density profile $n_i=n_i\Theta(x)$, 
where $\Theta(x)$ is the Heaviside step function).
In the electrostatic limit the phase matching with an 
impinging EM wave is possible 
only for particular target geometries 
where the frequency of electrostatic 
SWs depends on $k_{s}$ (e.g. a plasma slab of thickness $d$ where  
$\omega_{s}$ depends on the product $k_{s}d$ \cite{aliev}).
In our model, we assume a semi-infinite geometry but we do not use 
the electrostatic approximation, and
therefore in general the frequency $\omega_{s}$ of ESWs 
is significantly lower than $\omega_p/\sqrt{2}$ and depends 
on $k_{s}$, approaching the EM limit 
$\omega_{s} \simeq k_{s} c$ when $\omega_{s} \rightarrow 0$.
This allows the matching conditions (\ref{eq:MC}) to be 
satisfied in a semi-infinite plasma.
It is interesting to notice that in Ref.\cite{gradov2} 
it has been shown that two counterpropagating
ESWs generate nonlinear radiation at a frequency that is the
sum of the ESW frequencies; 
this process may be regarded, 
in some sense, as the inverse process of the TSWD.

One of the main results of this paper is that the 
intense ${\textbf v} \times {\textbf B}$ force at $2\omega$ 
of the laser pulse may drive a 
``$2\omega \rightarrow \omega+\omega$'' decay 
with a growth time of a few laser cycles.
In this paper, we focus on this case 
because of its direct relevance to the interpretation
of recent numerical simulations in two dimensions (2D) 
for normal laser incidence, which have provided 
evidence for this process \cite{macchi}.
Note that the surface oscillations observed in Ref.\cite{macchi} 
and theoretically studied in the present paper are different from 
the grating-like surface inhomogeneities induced by the magnetic 
force that were studied in Ref.\cite{plaja}; these latter 
``non-parametric'' surface structures oscillate at the same 
frequency $2\omega$ of the driving force and are generated at 
oblique incidence only. In the case of a $p$-polarized laser pulse 
at oblique incidence, we show below that the parametric process may 
be driven by the electric force leading to the generation of 
subharmonic ESWs around the frequency $\omega/2$.

The primary aim of this paper is to describe the ``basic 
principle'' of the TSWD as a possible route to the generation
of surface waves and to outline an analytical model for
the TSWD. Therefore, for the sake of simplicity, we use a cold 
fluid, non-relativistic plasma model that can be tackled
analytically. However, we believe that the TSWD is a general
process that may occur for laser and plasma parameters 
beyond the limits of the approximations used in the
present paper.  
Relativistic and kinetic effects, the
inclusion of damping mechanisms and the fully nonlinear
evolution of the TSWD are left for future investigations.

\section{The model}

In our model, we consider a cold
plasma with immobile ions and a step-like density 
profile $n_i(x)=n_i\Theta(x)$. 
The laser pulse has frequency $\omega$, 
linear polarization and wavevector 
${\textbf k}=(\omega/c)(\hat{\textbf x}\cos\theta+\hat{\textbf y}\sin\theta)$.
Translational invariance along $z$ is assumed. 
For both $s$- and $p$-polarizations, 
all fields oscillating at the frequency $\omega$
can be determined inside and outside the plasma 
from general Fresnel formulae for refraction and transmission
at a boundary \cite{jackson1}
between vacuum and a medium with an (imaginary) refractive index
${\sf n}$ given by ${\sf n}^2=1-\omega_p^2/\omega^2<0$, 
$\omega_{p}=(4\pi n_i e^2/m_e)^{1/2}$
being the plasma frequency.

In what follows we study the excitation of ``{\sf H}''
ESWs with the magnetic field  in the $z$ direction.
Thus, we deal with an effectively 2D geometry,
in which ${\textbf B}=B\hat {\textbf z}$ and the
Maxwell-Euler equations read
\begin{eqnarray}
\nabla\cdot{\textbf E}=4\pi e(n_0-n_e), 
\label{eq:poisson}\\
\nabla\times{\textbf E}=-\frac{1}{c}\partial_t B_z, 
\label{eq:stokes}\\
\nabla\times {\textbf B}=\frac{4\pi}{c}{\textbf j}+\frac{1}{c}\partial_t {\textbf E},
\label{eq:faraday}\\
m_e\partial_t {\textbf v}=
-m_e{\textbf v}\cdot\nabla {\textbf v}
-e\left({\textbf E}+\frac{\textbf v}{c}\times{\textbf B}\right)
\label{eq:euler}
\end{eqnarray}
and the current density is given by ${\textbf j}=-e n_e {\textbf v}$.

We adopt the following expansion for all fields
\begin{equation}
{F}(x,y,t)={F}_i(x)+\epsilon F_{0}(x,t-y\sin\theta/c)
                  +\epsilon^2[f_{+}(x,y,t)+f_{-}(x,y,t)],
\label{eq:expansion}
\end{equation}
where $\epsilon$ is a small expansion parameter and 
$f$ stands either for the electron density or the velocity
or for the EM fields in the $(x,y)$ plane.
In this expansion, $F_i$ represents unperturbed fields
of zero order (e.g., $n_i$).
The term $F_0$ represents the ``pump'' field 
at the frequency $\omega_0$,
that is taken to be of order $\epsilon$ and is written 
as 
\begin{equation}
F_{0}= \tilde{F}^{(\omega_0)}(x)e^{i\omega_0 (t-y\sin\theta/c)}
+\mbox{c.c.}
\end{equation}
In what follows the
cases with $\omega_0$=$\omega$ and $\omega_0=2\omega$ are 
studied separately.
The last term in (\ref{eq:expansion}) 
is the sum of two counterpropagating surface modes, 
which are assumed to be of order $\epsilon^2$ and are written as
\begin{equation}
f_{\pm}=\frac{1}{2}
\left[ \tilde{f}_{\pm}^{(\omega_{\pm})}(x)e^{ik_{\pm}y-i\omega_{\pm}t} 
                  \right] +\mbox{c.c.}.
\label{eq:envelopes}
\end{equation}
Using expansion (\ref{eq:expansion}), the coupling between
the pump and the surface modes is of order $\epsilon^3$.
Thus, from 
eqs.(\ref{eq:poisson}-\ref{eq:stokes}-\ref{eq:faraday}-\ref{eq:euler}) 
we obtain to order $\epsilon$ the pump fields. 
To order $\epsilon^2$, we obtain a set of linearized, homogeneous 
2D Maxwell-Euler equations which have ESWs as normal modes. 
Note that terms such as ${\textbf v}_{0}\cdot \nabla {\textbf v}_{0}$ 
in the Euler equation
are also of order $\epsilon^2$,
but are non resonant with the ESWs
at the frequencies $\omega_{\pm}$.
Such terms represent a source term for harmonic
oscillations at $2\omega_0$ and can be neglected
in our model.

ESWs are ``{\sf H}'' waves which can propagate in a dielectric medium
along a layer of discontinuity of the dielectric function, 
with the latter changing sign across the boundary \cite{landau}. 
For a cold, step-boundary plasma there is no volume 
charge density perturbation associated with SWs, i.e. 
$\nabla \cdot {\textbf E}=-4\pi e\delta n_e=0$. 
At a vacuum-plasma interface, 
electrons do not enter the vacuum side, but are ``stopped'' at the
interface ($x=0$) forming a surface charge layer. 

The dispersion relation for an SW of frequency $\omega_{s}$
and wavevector $k_s$ is
\begin{equation}
k_s^2=\frac{\omega_s^2}{c^2}
\frac{\omega_p^2-\omega_s^2}{\omega_p^2-2\omega_s^2}
=\frac{\omega_s^2}{c^2}
\frac{\alpha_s-1}{\alpha_s-2} 
.
\label{eq:SW_DR}
\end{equation}
We have set for convenience $\alpha_s=\omega_p^2/\omega_s^2$.
The dispersion relation (\ref{eq:SW_DR}) is shown 
in Fig.\ref{fig:ESW}.

The EM field envelopes (see eq.(\ref{eq:envelopes})) are given by   
\begin{eqnarray}
\tilde{E}^{s}_y(x)=\tilde{E}^{s}_y(0)\left[\Theta(-x)e^{q_{<}x}
+\Theta(x)e^{-q_{>}x}\right]  , \nonumber \\
\tilde{B}^{s}_z(x) 
= \frac{i\omega_s /c}{q_{<}}
\tilde{E}^{s}_y(0)\left[\Theta(-x)e^{q_{<}x}
 +\Theta(x)e^{-q_{>}x}\right]  ,
\nonumber \\
\tilde{E}^{s}_x 
= ik \tilde{E}^{s}_y(0) \left[\Theta(-x)\frac{e^{q_{<}x}}{q_{<}}
                              -\Theta(x)\frac{e^{-q_{>}x}}{q_{>}} \right] ,
\label{eq:SW_fields}
\end{eqnarray}
where
$q_{>}=({\omega_s}/{c})({\alpha_s-1}/{\sqrt{\alpha_s-2}})$,
$q_{<}=({\omega_s}/{c})({1}/{\sqrt{\alpha_s-2}})$.
The velocity fields are found from 
$\tilde{\textbf v}^{s}=(ie/m_e\omega_s){\textbf E}^{s}$.

All feedback effects of the nonlinear coupling on the 
pump fields are neglected. The coupling terms that lead
to the excitation of the ESW with frequency $\omega_{\pm}$ 
may be represented by the nonlinear force 
\begin{equation}
{\textbf f}^{(NL)}_{\pm}
=\tilde{\textbf f}^{(NL)}_{\pm}(x)e^{ik_{\pm}y-i\omega_{\pm}t}
=- \left[
       m_e ({\textbf v}_{\mp}\cdot \nabla {\textbf v}_{0}
       +{\textbf v}_{0}\cdot \nabla {\textbf v}_{\mp})
        +\frac{e}{c}({\textbf v}_{0}\times {\textbf B}_{\mp}
                    +{\textbf v}\times{\textbf B}_{0})\right]_{res}.
\label{eq:f_NL}
\end{equation}
In the term in square brackets on the r.h.s of (\ref{eq:f_NL}) 
only ``resonant'' terms with
the same phase of ${\textbf f}^{(NL)}_{\pm}$ are included.
Obviously, these terms exist if the matching conditions
(\ref{eq:MC}) hold.

The nonlinear coupling leads to the parametric excitation and
growth of ESWs.
We thus let the ESW envelopes vary slowly in time, i.e.
\begin{equation}
\tilde{f}_{\pm}(x) \rightarrow \tilde{f}_{\pm}(x,\epsilon t)
\end{equation}
To evaluate the growth rate, we use an energy approach.
The surface energy density 
of the ESW that will be needed in the calculations
is given by
\begin{equation}
U_{\pm}=\frac{k}{2\pi}\int_{-\pi/k}^{+\pi/k}\!dy\int_{-\infty}^{+\infty}\!dx
\left\langle u^{(K)}_{\pm}+u^{(F)}_{\pm}\right\rangle  ,
\end{equation}
where the brackets denote average over one laser cycle and $u^{(K)}_{\pm}$ and $u^{(F)}_{\pm}$ are the volume densities of the 
kinetic and EM fields energies, respectively:
\begin{eqnarray}
u^{(K)}_{\pm}&=&\frac{m_e n_0}{2} \left|{\textbf v}_{\pm}\right|^2  , \\
u^{(F)}_{\pm}&=&\frac{1}{8\pi}\left(\left|{\textbf E}_{\pm}\right|^2
                             +\left|{B}_{\pm,z}\right|^2\right)  .
\end{eqnarray}
The integral over $y$ yields a factor $1/2$.
The kinetic energy contribution vanishes for $x<0$. 
Using eqs.(\ref{eq:SW_fields}) and
$\tilde{v}_{\pm,x}(0^+)
=\tilde{v}_{\pm,y}(0)/\sqrt{\alpha-1}$,
and finally summing the two contributions 
we obtain for the total surface energy $U_{\pm}=M_{\pm}|\tilde{v}_{\pm,y}|^2/2$
where
\begin{eqnarray}
M_{\pm}=\frac{m_e n_i c}{4\omega_{\pm}}
\frac{\alpha_{\pm}(\alpha_{\pm}-2)^{1/2}(\alpha_{\pm}^2-2\alpha_{\pm}+2)}
     {(\alpha_{\pm}-1)^2} .
\label{eq:ESW_energy}
\end{eqnarray}

To obtain an expression for the temporal variation of 
$U_{\pm}$ due to the nonlinear coupling forces, 
we write the Euler and the Ampere equation for the ESW fields 
keeping terms up to order $\epsilon^3$ and disregarding terms 
that are non-resonant with the oscillation:
\begin{eqnarray}
m_e \partial_t {\textbf v}_{\pm}=- e {\textbf E}_{\pm}
+\epsilon {\textbf f}_{\pm}^{(NL)}  ,
\label{eq:GR_Euler} \\
\nabla\times{\textbf B}_{\pm}=\frac{4\pi}{c}
\left(-e n_o {\textbf v}_{\pm}+\epsilon {\textbf J}_{\pm}^{(NL)}\right)
+\frac{1}{c}\partial_t{\textbf E}_{\pm},
\label{eq:GR_current}
\end{eqnarray}
where ${\textbf f}_{\pm}^{(NL)}$ is given by eq.(\ref{eq:f_NL}),
and ${\textbf J}_{\pm}^{(NL)}=-e\delta n_e^{(\omega_0)} {\textbf v}_{\pm}$.
Using eqs.(\ref{eq:GR_Euler}-\ref{eq:GR_current}) together with Maxwell's
equations we obtain for the temporal variation of the energy densities
\begin{eqnarray}
\partial_t u^{(K)}_{\pm}= -e n_o {\textbf v}_{\pm}\cdot {\textbf E}_{\pm} 
                +\epsilon {\textbf f}_{\pm}^{(NL)}\cdot {\textbf v}_{\pm},\\
\partial_t u^{(F)}_{\pm}= -\nabla \cdot {\textbf S}_{\pm}
    -\epsilon {\textbf J}_{\pm}^{(NL)}\cdot {\textbf E}_{\pm},
\end{eqnarray}
where ${\textbf S}_{\pm}=(c/4\pi){\textbf E}_{\pm}\times{\textbf B}_{\pm}$
is the Poynting vector of the ESW. 
We integrate the sum of the energy densities over space and average over time.
Noting that the total flux of ${\textbf S}_{\pm}$
vanishes because of the evanescence
of the ESW fields, we finally obtain 
the equation for the evolution of the total energy $U$ 
of the ESW:
\begin{eqnarray}
\partial_t U_{\pm} &=&
\frac{2\pi}{k}\int_{-\pi/k}^{+\pi/k}\!dy\int_{0}^{+\infty}\!dx 
\partial_t\left\langle 
\frac{m_e n_0}{2}\left|{\textbf v}_{\pm}\right|^2
                 +\frac{1}{8\pi}
                  \left( |{\textbf E}_{\pm}|^2+|{\textbf B}_{\pm}|^2 \right)
                    \right\rangle \nonumber \\
 &=&\epsilon \frac{2\pi}{k}\int_{0}^{+\infty}\!dx
\int_{-\pi/k}^{+\pi/k}\!dy 
\left\langle {\textbf v}_{\pm}
  \cdot \left( e\delta n_e^{(\omega_0)} {\textbf E}_{\pm}
               +{\textbf f}_{\pm}^{NL} \right ) \right\rangle
 . \label{eq:dtU} \label{eq:energy_var}
\end{eqnarray}
The integral extends for $x>0$ only since there are no electrons
for $x<0$. 

From eq.(\ref{eq:energy_var}), 
performing the integral and neglecting high order terms
one obtains two coupled equations in the 
general form
\begin{equation}
\frac{M_{\pm}}{2} \partial_t |\tilde{v}_{\pm}|^2
   =A_0 \omega |\tilde{v}_{+}||\tilde{v}_{+}|S_{\pm} \sin\varphi,
\label{eq:M-R}
\end{equation}
where $A_0$ is the amplitude of the pump mode in dimensionless
units, $S_{\pm}$ are positive factors depending on $\omega_p/\omega$
and $\theta$, and $\varphi$ is obtained from the 
phase factors of the pump and ESW modes as 
$\varphi=\phi_{+}+\phi_{-}-\phi_{0}$.
From (\ref{eq:M-R}) one easily obtains
\begin{equation}
\partial_t^2 |\tilde{v}_{\pm}| =(A_0 \omega)^2
                                \left(\frac{S_{+}S_{-}}{M_{+}M_{-}}\right)
                                \sin^2\varphi |\tilde{v}_{\pm}| .
\end{equation}
Setting $\sin\varphi=1$ simply corresponds to finding the relative 
phase of the growing modes with respect to the phase of 
the pump mode.
The growth rate of ESWs is then
\begin{equation}
\Gamma =A_0 \omega \sqrt\frac{S_{+}S_{-}}{M_{+}M_{-}}.
\end{equation}
In the next section we use the analytical method outlined above to 
study TSWD in two cases of particular relevance:
TSWD driven by the ${\textbf v} \times {\textbf B}$ force for normal
laser incidence (which we name ``$2\omega \rightarrow \omega+\omega$''
decay), and TSWD driven by the electric force for
oblique laser incidence and $p$-polarization
(``$\omega \rightarrow \omega/2+\omega/2$'' decay).

\section{``$2\omega \rightarrow \omega+\omega$'' decay}
\label{sec:2w}
To discuss the ``$2\omega \rightarrow \omega+\omega$'' 
decay, we restrict ourselves 
for simplicity to the case of normal incidence,
as was done in the simulations reported in
Ref.\cite{macchi}. 
Thus, the ESW frequency $\omega_{\pm}=\omega$ and 
$k_{+}=-k_{-}$
The case of oblique incidence is indeed very similar, the main
difference being that the matching condition
$k_{+}+k_{-}=(2\omega/c)\sin\theta$ holds, causing a shift
$\delta\omega \neq 0$ between the two ESWs. 
It is found that the growth rate decreases monotonically 
with increasing incidence angle, since the pump force has a 
maximum at normal incidence \cite{macchi2}.
 
The laser wave can be represented by a single component of 
the vector potential, $A_z=A_z(x,t)$,
that for $x>0$ is given by
\begin{equation}
A_z(x,t)=A_z(x)\cos\omega t=A_z(0) e^{-x/l_s}\cos\omega t  ,
\label{eq:Az}
\end{equation}
where $l_s=c/(\omega_{p}^2-\omega^2)^{1/2}$ is the screening length. Imposing boundary conditions for the incident and 
reflected waves one finds
$A_z(0)=2A_i (\omega l_s/c)(1+\omega^2 l_s^2 /c^2)^{-1/2}$, where 
$A_i$ is the amplitude of the incident field.

Electrons perform
their quiver motion in the $z$ direction. Thus, 
the magnetic force term is in the $x$ direction and has a secular term
($0\omega$), named the ponderomotive force, 
and an oscillating term ($2\omega$) that, 
in what follows, we simply name 
the ${\textbf v}\times{\textbf B}$ force. The secular term 
corresponds to radiation pressure and creates a 
surface polarization of the plasma. The ${\textbf v}\times{\textbf B}$ force
drives a longitudinal, electrostatic oscillation
that acts as a pump mode for TSWD.
In the expansion (\ref{eq:expansion}), the pump amplitude 
is supposed to be of order $\epsilon$. 
As will be shown below, coupling between 1D and 2D fields
occurs only in the overdense plasma ($x>0$).
Thus, since the ${\textbf v}\times{\textbf B}$ force 
is quadratic in the laser field, the expansion 
(\ref{eq:expansion}) also implies $a(0) \sim \epsilon^{1/2}$, where
$a(0)$ is the (dimensionless) laser amplitude {\em at the surface} 
of the plasma, e.g 
$a(0)=\left({eA_z(0)}/{mc^2}\right)$.
For overdense plasmas
$a(0) \sim (\omega/\omega_p) a_i < a_i$, 
where $a_i=\left({eA_i}/{mc^2}\right)$. This yields
$\epsilon \sim \left({\omega}/{\omega_p}\right)^2 a_i^2 
=\left({n_c}/{n_e}\right) a_i^2 $,
where $n_c=m_e\omega^2/4\pi e$ is the ``critical'' density.
We therefore expect our expansion procedure to be 
valid even at relativistic fields amplitudes 
$a_i \geq 1$ for high enough plasma densities.

We now derive the solutions for the 1D pump fields at $2\omega$.
The electron oscillation velocity is obtained from the conservation
of canonical momentum as $m_e v_z=eA_z/c$.
For normal laser incidence,
the longitudinal ${\textbf v}\times{\textbf B}$ force is
given by
\begin{eqnarray}
-\frac{e}{c}v_z B_y &=&-\frac{e^2}{m_e c^2}A_z \partial_x A_z 
\equiv F^0(x)(1+\cos 2\omega t) ,
\label{eq:PM_force}
\end{eqnarray}
where we have set
$F^0 (x) =({m_e c^2}/{2 l_s }) a_s^2 e^{-2x/l_s}
\equiv F^0 e^{-2x/l_s}$.
To order $\epsilon$, one obtains the following equations
for the longitudinal, electrostatic motion :
\begin{eqnarray}
m_e \partial_t  V_x^{(2\omega)} 
= -e (E_x^{(0)}+E_x^{(2\omega)})
 + F^0 (x)(1+\cos 2\omega t)  , 
\label{eq:1DEuler}\\
\partial_x (E_x^{(0)}+E_x^{(2\omega)}) = 
-4\pi e (\delta n_e^{(0)}+\delta n_e^{(2\omega)})  , 
\label{eq:1DPoisson}\\
\partial_t \delta n_e^{(2\omega)}=-n_0 \partial_x V_x^{(2\omega)}. 
\label{eq:1Dcont}
\end{eqnarray}
All fields in the equations above 
decay inside the plasma as $\exp(-2x/l_s)$. 
The secular part simply gives $eE_x^{(0)}(x)=F^0(x)$  and 
$\delta n_e^{(0)}= -\partial_x F^0 (x)/4\pi e^2 = F^0 (x)/2\pi e^2 l_s$. 
For the motion at
$2\omega$ one obtains 
\begin{eqnarray}
\tilde{V}_x^{(2\omega)} &=& \frac{-i F^0}{\omega m_e D}
\left(\frac{\omega^2}{\omega_p^2}\right )e^{-2x/l_s}  , \label{eq:Vx2w} \\
\delta\tilde{n}_e^{(2\omega)} &=& n_i 
\frac{F^0}{\omega_p^2 l_s m_e D}e^{-2x/l_s}
 , \label{eq:ne2w} \\
e\tilde{E}_x^{(2\omega)} &=& \frac{F^0}{2D}e^{-2x/l_s} 
 . \label{eq:Ex2w}
\end{eqnarray}
The denominator $D$ is given by
\begin{equation}
D=1-\frac{4\omega^2}{\omega_p^2}=1-\frac{4n_c}{n_e}  ,
\end{equation}
which shows the well-known resonance at $n_e=4n_c$
due to excitation of plasmons with frequency $\omega_p=2\omega$ 
by the ${\textbf v}\times{\textbf B}$ force.

The difference between the total number of electrons and ions for $x>0$ is, 
to order $\epsilon$, 
\begin{eqnarray}
\Delta N_e^{(x>0)}=\int_0^{+\infty}\!dx \left(\delta{n}_e^{(2\omega)}+
\delta{n}_e^{(0)} \right)
=\frac{F^0}{2\pi e^2}
\left(1+\frac{\cos 2\omega t}{D} \right )  .
\end{eqnarray}
The fact that $\Delta N_e^{(x>0)}>0$ during most of the oscillation
implies that, due to compression 
from the ponderomotive and  ${\textbf v}\times{\textbf B}$ 
forces, electrons leave behind a charge 
depletion layer of thickness $\zeta=\Delta N_e^{(x>0)}/n_i$. 
We note that $\zeta$ is of order
$\epsilon$, thus to lowest order it is correct to treat the charge 
depletion layer as a surface layer. 
Heuristically, the oscillating behavior of $\zeta$ describes 
the plasma ``moving mirror'' effect \cite{bulanov}:
due to charge depletion the laser is reflected at $x=\zeta$ rather 
than exactly at $x=0$. This leads to the appearance of high 
harmonics in the reflected light.

Since $|D|<1$, there is always a time interval in which electrons 
are pulled into vacuum forming a cloud of negative charge. 
This interval is very short for $n_e/n_c \gg 4$, 
i.e. $D \simeq 1$, but for lower densities it is relevant  
in the interaction process. The motion in 
vacuum is strongly anharmonic and more difficult
to solve analitically than the motion inside the plasma. 
However, it is found in section \ref{sec:2w} that
the surface modes gain energy during the phase of electron motion
inside the plasma only, so that the 
expressions of fields for $x<0$ are not needed.
Nevertheless, we have to assume that the density of the 
electron cloud for $x<0$ is low enough for the 
laser propagation not to be affected.

Using eq.(\ref{eq:Vx2w}),
the NL force (\ref{eq:f_NL}) may be written in terms of the
oscillation velocity $V_x^{(2\omega)}$ as follows:
\begin{eqnarray}
{\textbf f}_{\pm}^{NL} &=&
- \left[m_e\left(V_x^{(2\omega)}(x) \partial_x {\textbf v}_{\mp} 
                   +v_{x,\mp}\partial_x V_x^{(2\omega)}(x) 
                    \hat{\textbf x} \right)
            -\frac{e}{c}V_x^{(2\omega)}(x)B_{z,\mp}\hat{\textbf y}
            \right]  .
\label{eq:f_w}
\end{eqnarray}              

Because of the symmetry with respect to inversion of the $y$ axis, 
the two ESWs may differ only by a
phase factor and have the same amplitude and energy.
Inserting the force (\ref{eq:f_w}) in the energy equation
(\ref{eq:energy_var})
and performing the integral in $y$ we obtain for the ESW energy
$U=U_{\pm}$
\begin{eqnarray}
\partial_t U
&=&\frac{1}{8}
\sum_{l=+k,-k}\int_{0}^{+\infty}\!dx 
\tilde{\textbf v}^{*}_{+l}
  \cdot \left (e\delta \tilde{n_e}^{(2\omega)}(x)\tilde{\textbf E}^{*}_{-l}
  \right.
-m_e n_0 \tilde{v}^{*}_{x,-l}\partial_x V_x^{(2\omega)}(x)
  \nonumber \\
       & & 
-m_e n_0 \tilde{V}_x^{(2\omega)}(x)\partial_x \tilde{\textbf v}^{*}_{-l}
+n_0 \frac{e}{c} \tilde{V}_x^{(2\omega)}(x)\tilde{B}^{*}_{z,-l}\hat{\textbf y}
\left. \right )   
+ \text{ c.c.}
\end{eqnarray}
We rewrite all fields as functions of $\tilde{V}_x^{(2\omega)}$
and $\tilde{\textbf v}_{\pm}$.
We obtain, after some algebra
\begin{eqnarray}
\partial_t U=\frac{m_e n_i}{2(1+q_{>} l_s)}
 \Re\left\{\tilde{V}_x^{(2\omega)}
\left[(1+q_{>}l_s)\tilde{\textbf v}^{*}_{+}\tilde{\textbf v}^*_{-}
      +2 \tilde{v}^{*}_{x,+}\tilde{v}^{*}_{x,-}
      -\frac{\tilde{v}^{*}_{y,+}\tilde{v}^{*}_{y,-}}{q_{>}l_s}\right],
\right\}.
\label{eq:nuU}
\end{eqnarray}

All the terms of 
eq.(\ref{eq:nuU}) are proportional to 
\begin{equation}
\Re\left(\tilde{V}_x^{(2\omega)}\tilde{v}^{*}_{+}\tilde{v}^{*}_{-}\right)
=\left|\tilde{V}_x^{(2\omega)} \tilde{v}_{\pm}^2\right|
\cos(\phi+\pi/2)  ,
\label{eq:Vv2}
\end{equation}
where $\phi=\phi_{+}+\phi{-}$, $\phi_{\pm}$ being the phase angles
of $v_{\pm}$, and we used eq.(\ref{eq:Vx2w}) and the fact 
that $F^{0}$ is real and positive. 
Thus, the phase of the growing ESWs with respect to
the ${\textbf v}\times{\textbf B}$ force is such that
$\phi=-\pi/2$, the value 
for which the growth rate is positive and has a maximum.
The overlap of the two ESWs gives
\begin{eqnarray}
\frac{\tilde{v}_{+}}{2}e^{iky-i\omega t} 
+\frac{\tilde{v}_{-}}{2} e^{-iky-i\omega t}
&=&\frac{|\tilde{v}_{\pm}|}{2}e^{-i\omega t}
\left(e^{iky+i\phi_+} +e^{-iky+i\phi_-}\right)  \nonumber \\
&=&|\tilde{v}_{\pm}| e^{-i\omega t-i\pi/4}
\cos\left(ky+\Delta\phi \right)  ,
\end{eqnarray}
where $\Delta\phi = (\phi_+ -\phi_-)/2$.
This is a standing wave which has a temporal phase
shift $-\pi/4$ with respect to the ${\textbf v}\times{\textbf B}$ force 
of eq.(\ref{eq:PM_force}). The phase is
such that, at a given position in 
$y$, the temporal maxima of ${V}_x^{(2\omega)}$ 
and ${v}^{(\omega)}$ overlap
once for laser cycle. 
The angle $\Delta\phi$ gives the location of the maxima of the 
standing wave on the $y$ axis, which
depends on the arbitrary choice of the 
initial phase. 

Using eq.(\ref{eq:Vv2}), we can rewrite the energy variation as
\begin{eqnarray}
\partial_t U =m_e n_i |\tilde{V}_x^{(2\omega)}|
\left[|\tilde{\textbf v}_{+}|^2
+\frac{2|\tilde{v}_{x,k}|^2}{1+q_{>}l_s}
-\frac{2|\tilde{v}_{y,k}|^2}{q_{>}l_s(1+q_{>}l_s)}
\right] 
\label{eq:nuU3}
 .
\end{eqnarray}
Note that $\partial_t U>0$ since 
$q_{>}l_s=\sqrt{(\alpha-1)(\alpha-2)^{-1}}>1$,
where $\alpha=\omega_p^2/\omega^2=n_e/n_c$.
Eliminating $\tilde{V}_x^{(2\omega)}$ as a function of 
$a_i$ and $\alpha$, after some algebra the growth rate
$\Gamma$ is obtained:
\begin{eqnarray}
\Gamma &=&{4\omega a_i^2}
\frac{(\alpha-1)^{3/2}}{\alpha|\alpha-4|[(\alpha-1)^2+1](\alpha-2)^{1/2}}
\left[1+\frac{2}{\alpha(1+q_{>}l_s)}
-\frac{2(\alpha-1)}{\alpha q_{>}l_s(1+q_{>}l_s)} \right]\nonumber \\
&\simeq& {4\omega a_i^2}
\frac{(\alpha-1)^{3/2}}{\alpha|\alpha-4|[(\alpha-1)^2+1](\alpha-2)^{1/2}}  .
\label{eq:rate1}
\end{eqnarray}
The leading contribution was highlighted in the last equality.

A plot of the growth rate as a function of 
$\alpha$ is given 
in Fig.\ref{fig:rate1}. For moderately overdense
plasmas, $\Gamma$ is a considerable fraction of the 
laser frequency.
The resonance at $n_e=4n_c$ is due to the resonant 
behavior of the pump field.
The growth rate diverges also in the electrostatic limit
when $n_e \rightarrow 2n_c$. 
However, in this limit the value of $k$ tends to infinity, i.e. the 
wavelength becomes very small and one expects that this second 
resonance might be damped by thermal effects, 
which are neglected in our model and are left for future
inevstigations.
 
The comparison of spatial and temporal 
scales predicted from this model
with the simulation results shows reasonable agreement for 
laser and plasma parameters such that our expansion 
procedure is valid \cite{macchi}. We note that simulation 
results for high intensities in the relativistic regime 
suggest that a $2\omega \rightarrow \omega+\omega$ TSWD-like
process still occurs and produces strong rippling of the plasma
surface; the spatial scales are 
different from those predicted by our analytical,
cold fluid model, but still of the same order of magnitude.

We conclude this section by noting that, although we took
the laser polarization along $z$, the whole derivation
is valid at normal incidence for any polarization 
direction. The only difference in 2D geometry is that,
for the laser polarization along $y$, the quiver oscillations
at $\omega$ and the ESW oscillations at the same frequency
overlap. This is observed in numerical simulations \cite{macchi2}.

\section{``$\omega \rightarrow \omega/2+\omega/2$'' decay}
\label{sec:w}
A case of particular interest is that of oblique incidence
and $p$-polarization of the laser pulse, since these 
conditions lead in general to a stronger plasma coupling.
In the preceding section we 
have found that the $2\omega \rightarrow \omega+\omega$
decay is driven by the longitudinal component of the 
velocity field, as shown by eq.(\ref{eq:f_w}).
 For $p$-polarization and $\theta \neq 0$ 
the electric field drives a strong longitudinal oscillation
at the frequency $\omega$; thus,
one expects that the TSWD process leads to 
the generation of {\em subharmonic} ESWs, having 
frequencies around $\omega/2$,
with peak efficiency at some angle of incidence.
As shown below, this picture is substantially correct.
However, there is a small but non-vanishing 
growth rate for the 
$\omega \rightarrow \omega/2 + \omega/2$ process
even at normal incidence. In fact, the 
pump force (\ref{eq:f_NL}) is now given by
\begin{eqnarray}
{\textbf f}_{\pm}^{(NL)}
&=&- m_e (v_{\mp,x}\partial_x {\textbf V}^{(\omega)}
        +V^{(\omega)}_x\partial_x {\textbf v}_{\mp})
        -m_e (v_{\mp,y}\partial_y {\textbf V}^{(\omega)}
        +V^{(\omega)}_y\partial_y {\textbf v}_{\mp}) \nonumber \\
     & &  -\frac{e}{c}({\textbf V}^{(\omega)}\times {\textbf B}_{\mp}
        +{\textbf v}_{\mp}\times{\textbf B}^{(\omega)}),
\end{eqnarray}
and does not vanish for $\theta=0$. 

The case of oblique incidence and 
$p$-polarization of the laser pulse
may be tackled with an approach analogous
to that of section \ref{sec:2w}.
The main physical difference with the
previous case is that now the ``pump'' is a divergence-free 
velocity field at the laser frequency $\omega$,
and its amplitude is now proportional to the laser
field rather than to the laser intensity.
In applying the expansion procedure (\ref{eq:expansion}) again,
now $\epsilon \sim (\omega/\omega_p)a_i$ 
(appendix \ref{app:w/2}).
Thus, the limits
of validity of this expansion are more restrictive in the present
case: the expansion tends to be valid only for rather high densities 
or rather low fields. One must also note that 
temperature effects, neglected in the cold plasma approximation,
are important for low fields.

The calculation of the growth rate for subharmonic ESWs 
can be performed by the same method as in the preceding sections 
in a straightforward way. The details are reported
in appendix \ref{app:w/2}. As a final result, the growth rate 
$\Gamma=\Gamma(\alpha,\theta)$ may be written as 
\begin{eqnarray}
\Gamma  = a_i \omega |{\cal F}_B|\left(\frac{n_c}{n_e}\right)^{5/2}
                   K(\alpha,\theta),
\label{eq:rate2}
\end{eqnarray}
where ${\cal F}_B$ is the Fresnel factor for the magnetic field
and the factor $K(\alpha,\theta)$ scales weakly with density.
A plot of $\Gamma$ as a function of $\theta$ for different values of
$\alpha$ is reported in Fig.\ref{fig:rate2}. 

The frequency shift $\delta\omega$ of the two 
ESWs can be calculated as a function of $\theta$ and
$\alpha$ from the matching condition
$k_{+}+k_{-}=(\omega/c)\sin\theta$. 
The result is shown in Fig.\ref{fig:shift}.
Comparing with Fig.\ref{fig:rate2}, one finds 
that $\delta\omega \approx 0.1 \omega$ in 
conditions  favorable to the 
$\omega \rightarrow \omega/2+\omega/2$
TSWD.

\section{Discussion}
The TSWD concept has been investigated analytically in two cases
that clarify its basic features in the context of the cold 
fluid plasma approximation. 
This process may lead to the excitation of ESWs by an ultrashort 
laser pulse in a solid target without any special (e.g., grating-like) 
structure, and even for normal incidence and $s$-polarization.

In both the cases discussed in this paper,
we found a strong decrease of the TSWD growth rate 
for increasing densities. This does not necessarily imply 
that TSWD is not relevant to short pulse interaction with solid 
targets which have $n_e/n_c \gg 1$. 
In fact, important processes such as 
high harmonic generation or fast electron production 
are more efficient when the laser pulse interacts with 
a moderate density plasma. This is the case for most of
the experiments of short pulse interaction with solid targets,
since a moderate density ``shelf'' is  usually produced
at the time of peak pulse intensity
by target ablation and plasma 
hydrodynamic expansion during the leading edge 
of the short pulse or during the long prepulse. 
For instance, the ``$2\omega \rightarrow \omega + \omega$'' 
decay appears to be very efficient exactly in conditions favorable for
high harmonics production via the ``moving mirror'' effect
\cite{bulanov,gibbon}. Simulations \cite{macchi} suggest
that TSWD in strongly nonlinear and relativistic regimes 
(beyond the limits of the analytical approach of this paper) 
may produce surface perturbations acting as a ''seed'' for,
e. g. , electron instabilities leading to current filamentation  
\cite{califano,sentoku} or detrimental distortions of the moving 
mirrors, and for Rayleigh-Taylor-like (RT) hydrodynamic instabilities 
occuring on the time scale of ion motion. In Ref.\cite{macchi} 
it is estimated that the growth rate of the RT instability is much 
slower than that of TSWD. A different hydrodynamic instability with a 
surprisingly high growth rate has been reported in Ref.\cite{gamaly2}).

The $2\omega \rightarrow \omega + \omega$  and the 
$\omega \rightarrow \omega/2 + \omega/2$ 
processes have been considered
independently. This is appropriate since these are resonant 
processes which do not interfere with each other. 
In principle, both decays may occur during the interaction
of an intense laser pulse with an overdense plasma.
According to our analysis the 
$2\omega \rightarrow \omega + \omega$
process appears to have a stronger growth rate.
However, in the case of the 
$\omega \rightarrow \omega/2+\omega/2$
process our analytical approach is valid for a very narrow
range of parameters only. Either an extension of 
the present analytical approach or numerical simulations
are needed to investigate the TSWD for a very 
intense, $p$-polarized laser pulse at oblique incidence.

\section{Conclusions}
We have discussed the Two-Surface Wave Decay as 
a novel mechanism for the excitation 
of electron surface waves in the interaction of ultrashort,
intense laser pulses with overdense plasmas. TSWD
is based on the parametric excitation of a pair of 
ESWs. The ``pump'' force may either be the 
magnetic or the electric force of the laser pulse, and leads 
respectively to
the generation of two ESWs with frequencies close to 
the laser frequency of half of it. 
An analytical model for TSWD in the cold fluid plasma,
non-relativistic approximation has been developed.
The model supports the interpretation of recent simulation results
\cite{macchi} and suggests that TSWD may be of relevance in 
certain regimes of laser interaction with solid targets.

\acknowledgments
This work was partly supported by INFM trhough a PAIS project. 
We are grateful to Prof. L. Stenflo for bringing many references 
to our attention and to M. Battaglini for very useful discussions.

\appendix

\section{Growth rate of the 
$\omega \rightarrow \omega/2+\omega/2$ decay}
\label{app:w/2}
We now give the detailed derivation of the growth rate
for the $\omega \rightarrow \omega/2+\omega/2$ TSWD
of section \ref{sec:w}.
The pump fields are easily found 
with the help
of Fresnel formulae and Maxwell equations. For instance,
the magnetic field is given by
\begin{eqnarray}
B^{(\omega)}_z(x,t)&=&B^{(\omega)}_z(0^+)e^{iky\sin\theta-x/l_p-i\omega t}
+\mbox{ c.c. } ,
\end{eqnarray}
where the screening length is given by
\begin{equation}
l_p=\frac{c}{\omega_p}\left(1-\frac{\omega^2}{\omega_p^2}\cos^2\theta
\right)^{-1/2} =\frac{c}{\omega}\frac{1}{\sqrt{\alpha-\cos^2\theta}} ,
\end{equation}
and the magnetic field at the surface is given by the Fresnel formula
\begin{eqnarray}
\frac{B_{z}(0^+)}{B_{z,i}}&=&\frac{2{\sf n}^2 \cos\theta}
                   {\sqrt{{\sf n}^2-\sin^2\theta}+{\sf n}^2\cos\theta}
\nonumber \\
&=&\frac{2(\alpha-1) \cos\theta}
                   {(\alpha-1)\cos\theta-i\sqrt{\alpha-\cos^2\theta}}
\equiv {\cal F}_B(\theta),
\end{eqnarray}
where $B_{z,i}$ is the incident field amplitude in vacuum.
The electric field components are found by using 
$\nabla \cdot {\textbf E}^{(\omega)}=0$ and 
$\nabla \times {\textbf E}^{(\omega)}=i(\omega/c)B^{(\omega)}_{z}+\mbox{c.c.}$.  

The nonlinear force is given by eq.(\ref{eq:f_w}).
Evaluating the spatial derivatives as 
$\partial_x {\textbf V}^{(\omega)}=-{\textbf V}^{(\omega)}/l_p$, 
$\partial_y {\textbf V}^{(\omega)}=ik_{t}{\textbf V}$,
$\partial_y {\textbf v}_{\pm}=ik_{\pm}{\textbf v}_{\pm}$, 
and $\partial_x {\textbf v}_{\pm}=-q_{\pm}{\textbf v}_{\pm}$ 
where $q_{\pm} \equiv q_{>}(\omega_{\pm})$,
and keeping resonant terms only we find
\begin{eqnarray}
\tilde{f}_{\pm,x}^{(NL)}=\frac{m_e}{4}
    & &\left[
    \left(\frac{1}{l_p}+q_{\mp}\right)\tilde{V}^{(\omega)}_x \tilde{v}^*_{\mp,x}
    -\frac{e}{m_e c}\left(\tilde{V}^{(\omega)}_y \tilde{B}^*_{\mp,z}
                         +\tilde{B}^{(\omega)}_z \tilde{v}^*_{\mp,y}\right)
    \right.\nonumber \\ & & \left.
    -ik_{t} \tilde{V}^{(\omega)}_x \tilde{v}^*_{\mp,y}
    +ik_{\mp} \tilde{V}^{(\omega)}_y \tilde{v}^*_{\mp,x}
\right]e^{-(q_{\mp }+1/l_p)x} +\mbox{ c.c.},\\
\tilde{f}_{\pm,y}^{(NL)}=\frac{m_e}{4}
      & &\left[
          \frac{1}{l_p}\tilde{V}^{(\omega)}_y \tilde{v}^*_{\mp,x}
         +q_{\mp} \tilde{V}^{(\omega)}_x \tilde{v}^*_{\mp,y}
         +\frac{e}{m_e c}\left(\tilde{V}^{(\omega)}_x \tilde{B}^*_{\mp,z}
                         +\tilde{B}^{(\omega)}_z \tilde{v}^*_{\mp,x}\right)
    \right.\nonumber \\ & & \left.
         -i(k_t-k_{\mp})\tilde{V}^{(\omega)}_y\tilde{v}^*_{\mp,y} 
\right]e^{-(q_{\mp }+1/l_p)x} +\mbox{c.c.}.
\end{eqnarray}
We rewrite all ``pump'' terms as a function of $\tilde{V}^{(\omega)}_y$ 
by using $\tilde{V}^{(\omega)}_x=i k_{t} l_p \tilde{V}^{(\omega)}_y$,
$\tilde{B}^{(\omega)}_z=
({m_e c}/{e l_p})(k_t^2 l_p^2-1)\tilde{V}^{(\omega)}_y$.
In a second step we rewrite all SW fields as a function of 
$\tilde{v}_{\pm,y}$ by using 
$\tilde{v}_{\pm,x}=(ik_{\pm}/q_{\pm})\tilde{v}_{\pm,y}$
and  $\tilde{b}_{\pm,z}=
-m_e \omega_{\pm}^2 (\alpha_{\pm}-1)\tilde{v}_{\pm,y}/(ecq_{\pm})$.
We thus obtain 
\begin{eqnarray}
\tilde{f}_{\pm,x}^{(NL)}&=&\frac{m_e}{4}\tilde{V}^{(\omega)}_y\tilde{v}^*_{\mp,y}
      \left[ 
         \frac{k_{\mp}}{q_{\mp}}\left(k_t+k_t q_{\mp}l_p+k_{\mp}\right)
        +\frac{\omega_{\mp}^2}{q_{\mp} c^2}(\alpha_{\mp}-1)
        +\frac{1}{l_p}
                      \right]e^{-(q_{\mp}+1/l_p)x} \\
        &\equiv&\frac{m_e}{4}\tilde{V}^{(\omega)}_y\tilde{v}^*_{\mp,y}Q_{\pm,x}        
           e^{-(q_{\mp}+1/l_p)x} \\
\tilde{f}_{\pm,y}^{(NL)}&=&\frac{m_e}{4}\tilde{V}^{(\omega)}_y\tilde{v}^*_{\mp,y}
         i\left[ 
          k_t l_p \left(q_{\mp}-\frac{1}{l_p}
                        -\frac{k_t k_{\mp}}{q_{\mp}}
                        - \frac{\omega_{\mp}^2}{q_{\mp} c^2}(\alpha_{\mp}-1)
                   \right)
          +k_{\mp}
                  \right]e^{-(q_{\mp}+1/l_p)x}\\
         &\equiv&\frac{m_e}{4}\tilde{V}^{(\omega)}_y\tilde{v}^*_{\mp,y}iQ_{\pm,y}
            e^{-(q_{\mp}+1/l_p)x}.
\end{eqnarray}
The relations 
$\omega_{\pm}^2(\alpha_{\pm}-1)/q_{\pm}=\omega_{\pm} c \sqrt{\alpha_{\pm}-2}$,
$k_{\pm}/q_{\pm}=1/\sqrt{\alpha_{\pm}-1}$, 
$k_t l_p=\sin\theta(\alpha-\cos^2\theta)^{-1/2}$ may be used
to simplify the expressions for $Q_{\pm,x}$ and $Q_{\pm,y}$.
We thus obtain
\begin{eqnarray}
Q_{\pm,x}=\frac{1}{\sqrt{\alpha_{\mp}-1}}[k_t(1+q_{\mp}l_p)+k_{\mp}]
          +\frac{\omega_{\mp}}{c}\sqrt{\alpha_{\mp}-2}+\frac{1}{l_p},\\
Q_{\pm,y}=k_t l_p \left(q_{\mp}-\frac{1}{l_p}
                        -\frac{k_t}{\sqrt{\alpha_{\mp}-1}}
                        -\frac{\omega_{\mp}}{c}\sqrt{\alpha_{\mp}-2}\right)
                        +k_{\mp}.
\end{eqnarray}

In the energy equation (\ref{eq:energy_var}), since
$\delta n^{(\omega)}=0$ the integrand,
eliminating $\tilde{v}_{\pm,x}$,
is given by
\begin{eqnarray}
\left\langle {\textbf v}_{\pm}\cdot{\textbf f}_{\pm}^{NL}  \right\rangle
&=&\frac{m_e}{16}i\tilde{V}^{(\omega)}_y\tilde{v}^*_{\mp,y}\tilde{v}^*_{\pm,y}
               \left(\frac{k_{\pm}}{q_{\pm}}Q_{\pm,x}+Q_{\pm,y}\right)
               +\mbox{ c.c.}\\
&=&\frac{m_e}{8}
   \Re\left[i\tilde{V}^{(\omega)}_y\tilde{v}^*_{\mp,y}\tilde{v}^*_{\pm,y}\right]
   \left(\frac{k_{\pm}}{q_{\pm}}Q_{\pm,x}+Q_{\pm,y}\right).
\end{eqnarray}
We thus find
\begin{eqnarray}
\partial_t U_{\pm} =n_i \frac{m_e}{8}\frac{l_p}{1+({q_{+}+q_{-}})l_p}
                   \left(\frac{k_{\pm}}{q_{\pm}}Q_{\pm,x}+Q_{\pm,y}\right)
                   \left|\tilde{V}^{(\omega)}_y\tilde{v}_{\mp,y}\tilde{v}_{\pm,y}\right|
                   \sin\varphi
               .
\end{eqnarray}
Here, $\varphi=(\phi_{+}+\phi_{-}-\Phi)$, where $\Phi$ and $\phi_{\pm}$ are
the phase factors of $\tilde{V}^{(\omega)}_y$ and $\tilde{v}_{\pm,y}$, respectively.
It is worth rewriting these equations in a more compact form.
Introducing $a_i=eB_i/m_e\omega c$ and using Fresnel's formulas  
to eliminate $\tilde{V}_y$ we obtain
\begin{eqnarray}
\partial_t U_{\pm} =n_i \frac{m_e}{8} a_i c |{\cal F}_B(\theta)|
                   |\tilde{v}_{+,y}||\tilde{v}_{-,y}|
                   \sin\varphi G_{\pm}(\alpha,\theta),
\end{eqnarray}
where we have posed
$G_{\pm}(\alpha,\theta)=G_0(\alpha,\theta)g_{\pm}(\alpha,\theta)$             
and
\begin{eqnarray}
G_0=\frac{\sqrt{\alpha-\cos^2\theta}}{(\alpha-1)[1+({q_{+}+q_{-}})l_p]},
\end{eqnarray}
\begin{eqnarray}
g_{\pm}&=&l_p\left(\frac{k_{\pm}}{q_{\pm}}Q_{\pm,x}+Q_{\pm,y}\right)
   \nonumber \\
       &=&\frac{1}{\sqrt{(\alpha_{+}-1)(\alpha_{-}-1)}}
         [k_t l_p (1+q_{\mp}l_p)+k_{\mp}l_p]
        +\frac{\omega_{\mp}l_p}{c}
         \frac{\sqrt{\alpha_{\mp}-2}}{\sqrt{\alpha_{\pm}-1}}
        +\frac{1}{\sqrt{\alpha_{\pm}-1}} \nonumber \\ & &
        +k_t l_p \left(q_{\mp}l_p-1-\frac{k_t l_p}{\sqrt{\alpha_{\mp}-1}}
        -\frac{\omega_{\mp}l_p}{c}\sqrt{\alpha_{\mp}-2}\right)
                        +k_{\mp}l_p
.
\end{eqnarray}
Using the equations above we obtain the following coupled
equations in the form (\ref{eq:M-R}) for the field amplitudes
\begin{eqnarray}
\mu_{\pm}\partial_t |\tilde{v}_{\pm,y}|^2
=a_i \omega |\tilde{v}_{+y}| |\tilde{v}_{-y}| 
            |{\cal F}_B| G_{\pm}\sin\varphi ,
\end{eqnarray}
where $\mu_{\pm} \equiv 4M_{\pm}/{(m_e n_i c)}$.
Thus, the growing modes have relative phases such that 
$\sin\varphi=1$. The growth rate is given by
\begin{eqnarray}
\Gamma = a_i \omega |{\cal F}_B| G_0
                   \sqrt{\frac{|g_{+}g_{-}|}{\mu_{+}\mu_{-}}}
    \equiv a_i \omega |{\cal F}_B|\left(\frac{n_c}{n_e}\right)^{5/2}
                    K(\alpha,\theta)
\end{eqnarray}
which is the formula (\ref{eq:rate2}). A plot of $\Gamma$ is depicted
in Fig.\ref{fig:rate2}.

\begin{figure}
%\centerline{\includegraphics[width=8cm]{./Fig1_macchi.eps}}
\caption{Dispersion relation of electron surface waves
(thick lines) 
eq.(\ref{eq:SW_DR}), and the matching conditions for TSWD, 
eq.(\ref{eq:MC}).}
\label{fig:ESW}
\end{figure}

\begin{figure}
%\centerline{\includegraphics[width=8cm]{./Fig2_macchi.eps}}
\caption{The TSWD growth rate $\Gamma$ 
(normalized to $a_i^2 \omega$) 
for the 
$2\omega \rightarrow \omega+\omega$ 
process at normal
incidence, eq.(\ref{eq:rate1}), as a function of 
$\alpha=n_e/n_c$.}
\label{fig:rate1}
\end{figure}

\begin{figure}
%\centerline{\includegraphics[width=8cm]{./Fig3_macchi.eps}}
\caption{The TSWD growth rate $\Gamma$
(normalized to $a_i \omega$) 
for the $\omega \rightarrow \omega/2+\omega/2$ 
process at oblique incidence, eq.(\ref{eq:rate2}), 
as a function of $\theta$ and $\alpha=n_e/n_c$.}
\label{fig:rate2}
\end{figure}

\begin{figure}
%\centerline{\includegraphics[width=8cm]{./Fig4_macchi.eps}}
\caption{The frequency shift $\delta\omega$ 
for the $\omega \rightarrow \omega/2+\omega/2$ 
process at oblique incidence,
as a function of $\theta$ and $\alpha=n_e/n_c$.
The dashed line gives the shift  
$\delta\omega^{max}=(1/2)\sin\theta$ 
that is obtained for $\alpha \rightarrow \infty$.}
\label{fig:shift}
\end{figure}

%\newpage
%\centerline{\includegraphics[width=8cm]{./Fig1_macchi.eps}}
%\vfil {\bf Fig.1: A. Macchi et al., ``Two-surface wave decay''}

%\newpage
%\centerline{\includegraphics[width=8cm]{./Fig2_macchi.eps}}
%\vfil {\bf Fig.2: A. Macchi et al., ``Two-surface wave decay''}

%\newpage
%\centerline{\includegraphics[width=8cm]{./Fig3_macchi.eps}}
%\vfil {\bf Fig.3: A. Macchi et al., ``Two-surface wave decay''}

%\newpage
%\centerline{\includegraphics[width=8cm]{./Fig4_macchi.eps}}
%\vfil {\bf Fig.4: A. Macchi et al., ``Two-surface wave decay''}

\end{document}